# Speech Emotion Recognition Considering Local Dynamic Features

Haotian Guan[1], Zhilei Liu[1,*], Longbiao Wang[1], Jianwu Dang[1,2,*], Ruiguo Yu[1]

[1]Tianjin Key Lab. of Cognitive Computing and Application, Tianjin University, Tianjin, China
[2]Japan Advanced Institute of Science and Technology, Ishikawa, Japan

```
{htguan, zhileiliu, longbiao_wang, dangjianwu,
              rgyu}@tju.edu.cn
```

**Abstract.** Recently, increasing attention has been directed to the study of the speech emotion recognition, in which global acoustic features of an utterance are mostly used to eliminate the content differences. However, the expression of speech emotion is a dynamic process, which is reflected through dynamic durations, energies, and some other prosodic information when one speaks. In this paper, a novel local dynamic pitch probability distribution feature, which is obtained by drawing the histogram, is proposed to improve the accuracy of speech emotion recognition. Compared with most of the previous works using global features, the proposed method takes advantage of the local dynamic information conveyed by the emotional speech. Several experiments on Berlin Database of Emotional Speech are conducted to verify the effectiveness of the proposed method. The experimental results demonstrate that the local dynamic information obtained with the proposed method is more effective for speech emotion recognition than the traditional global features.

## 1       Introduction

As is all known, speech conveys some additional messages beyond the words, such as emotion or identity of the speaker. With the rapid development of human-computer interaction in recent years, there is a growing interest in the emotion recognition from speech. Recognizing the emotion from the speech is beneficial for machines to communicate with the human. However, this is a problem full of challenges because the expression of emotion varies from one person to another [1].

   The traditional method of speech emotion recognition is as follows. In most existing approaches, low-level features of each frame in an utterance are extracted firstly. Then, the statistical features such as mean, maximum, and minimum values of these frames are calculated from the whole utterance. However, taking the features of the whole emotional utterance into account is somewhat unreasonable since human's perception of emotional speech is diverse. Considering the arithmetic capability of computers, some salient features are usually selected to represent the natures of the emotional speech. Therefore, feature selection is critical to explore features which are more effective for the expression of emotion in order to improve the recognition performance. Finally, these selected salient features are fed into a classifier to conduct the speech emotion classification.

In most of the previous works, global acoustic features of an utterance are usually adopted to eliminate the content differences and reduce the number of features [2]. However, emotional information of the speech is usually characterized by its dynamic changes [3]. In other words, the emotion-related components varies with time, rather than being constant in an utterance. Thus, the utilization of global statistical features alone, which takes statistics of the features in a whole utterance, may disregard some local dynamic information of emotion in speech.

To take such local information into consideration, segmentation is simply used to avoid the shortcomings of global features. There have been some studies working on the segmentation of utterances for the classification of speech emotion. In the work of Björn Schuller et al. [4], several segmentation schemes are proposed. The experimental results show that the combination of global and relative time interval features makes a significant improvement. Je Hun Jeon et al. [5] compare different segment units (3-words, phrases, and time-based segment) and find that using time-based subsentence segment units outperforms others. Hao Zhang et al. [6] use different segment selection approaches based on entropy, mutual information, and correlation coefficients, which yields better performances. Krothapalli Sreenivasa Rao et al. [7] report that the performance due to local prosodic features is above that of global ones. All these previous research reveals the effectiveness of segmental features on speech emotion recognition compared with the global utterance features.

Besides, the prosodic features conveying significant emotional information are utilized and analyzed in many previous studies [2], [8]. Pitch, as one of the prosodic features, has been found discriminative across different emotions, to some extent. For example, the average pitch of the speech with anger or happiness emotion is usually higher than which with sadness or fear emotion. In addition, the contour of pitch also differs among the utterances with different emotions [9].

At the aspect of classifiers utilized in previous research, some unsupervised learning methods are commonly used, such as Gaussian Mixture Model (GMM) in [10]. In addition, Support Vector Machine (SVM), which is a kind of supervised learning method, is employed more because of its capability and performance for modeling small-scale data with fewer parameters to be trained. Its target is to find a hyperplane to distinguish the data. Recently, with the development of deep learning methods, Deep Neural Network (DNN), Deep Belief Network (DBN) and some other deep learning methods, which are based on the perception mechanism of the human brain, are also utilized in speech emotion recognition [11], [12]. However, largescale datasets are necessary for the training of such kind of deep learning methods.

In this paper, time-based segmentation approach, which is to divide an utterance according to the time without taking the lexical information, is utlized to capture the temporal information of the emotional speech. The utilization of time-based segmentation achieves higher real-time capability, which can improve the audio stream processing performance to certain degree. And a novel pitch probability distribution, which is obtained by drawing the histogram, is proposed as a local dynamic prosodic feature, since pitch plays an important role in the expression of emotion and the histogram can reflect the distribution of the values to a certain degree. Firstly, the pitch histogram and other acoustic features are extracted from each segment of the utterance. After that, an optional processing of principal components analysis (PCA) is adopted for feature selection. Finally, these selected features are fed into an SVM classifier and the predicted class of emotion are obtained. The proposed framework for speech emotion recognition is illustrated in Fig. 1. Several comparative experiments are designed to validate the effectiveness of the proposed method. Based on the comparison of the experimental results, we can conclude that the combination of segmentation and the pitch probability distribution features, which considers the local dynamic information, achieves better results.

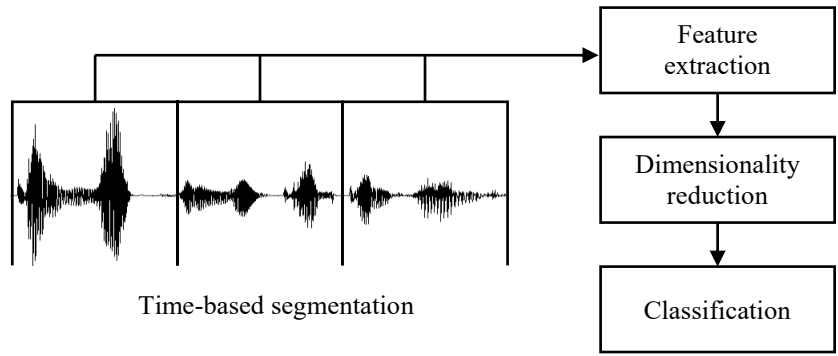

**Fig. 1.** Overview of the proposed method

The rest of this paper is organized as follows. In Section 2, the detailed method is provided, in which the time-based segmentation and the proposed novel pitch probability distribution features obtained by drawing the histogram are introduced. Experimental conditions and results are presented in Section 3. Discussion and conclusion are given in Section 4.

## 2 Time-based segmentation and local dynamic pitch probability distribution feature extraction

### 2.1 Time-based segmentation

The Relative Time Intervals (RTI) approach [4] is utilized for time-based speech segmentation. In addition, traditional Global Time Intervals (GTI) approach is adopted for comparison, which simply means using the whole utterance without segmentation and is usually used in traditionalmethods. Figure 2 shows the illustration of applying GTI and RTI approaches for the utterances with different lengths, in which the strips represent the utterances and the numbers refer to the indexes of the segments.

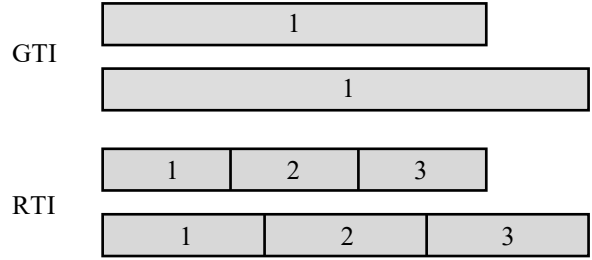

**Fig. 2.** Global Time Intervals (GTI) and Relative Time Intervals (RTI) approaches applied to a short and a long utterance respectively [4]

In the time-based segmentation approaches, taking the average is one of the simplest techniques to divide an utterance, which also guarantees the same number of the divided segments in an utterance. Therefore, in the RTI approach, as shown in Fig. 3, an utterance is first divided into n segments with the same duration of $\tau / n$, in which $\tau$ denotes the length of the utterance, and $n$ keeps invariant in the whole process. Next, each segment will be divided into frames of 25 ms length with 15 ms overlap.

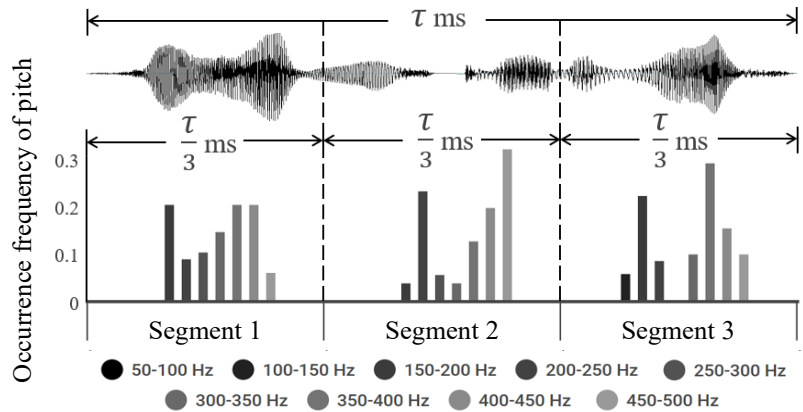

**Fig. 3.** Local dynamic pitch probability distribution feature extraction ($n = 3$)

## 2.2 Local dynamic pitch probability distribution feature extraction

After the segmentation, the pitch value of each frame is calculated, and only the values within a certain range are taken into account for pitch histogram computation. As shown in Fig. 3, the horizontal axis corresponds to several bins (or intervals) of the pitch range, while the vertical axis is the occurrence frequency of the pitch falling into each bin. The pitch histogram is normalized, with the sum of the heights equaling one.

When the range of pitch is set to [a, b] and the bin width is $h$, there will be $(b-a)/h$ bins for each segment in the histogram.

Finally, the value of each bin is concatenated and treated as the pitch probability distribution feature, and then is fed into the classifier for emotion recognition, together with some other features extracted from each segment, which are described in the next section. Z-score normalization is used to eliminate the difference in the scales of different kinds of features. The calculation is as follows:

$$x' = \frac{x - \mu}{\sigma} \quad (1)$$

where $\mu$ and $\sigma$ are the mean value and standard deviation of the population, respectively.

## 3     Experiments

### 3.1     Experimental conditions

In this paper, our proposed approach is experimentally evaluated on the commonly used Berlin Database of Emotional Speech (Emo-DB), which contains 535 utterances in German covering seven emotions [13]. Ten sentences without emotional content are acted by five actress and five actors, who are all professional. The distributions of the samples with different emotions in the database are 23.7% anger, 15.1% boredom, 8.6% disgust, 12.9% fear, 13.3% happiness, 11.6% sadness, and 14.8% neutral. 84.3% accuracy is reported for a human perception test.

The parameters for above mentioned time-based segmentation and feature extraction are set as following: $n = 3$, $a = 50$ Hz, $b = 500$ Hz, and $h = 50$ Hz. The pitch values are extracted with Praat [14].

Table 1. INTERSPEECH Emotion Challenge 2009 feature set

| LLDs (16×2) | Functionals (12) |
| --- | --- |
| (Δ)ZCR, (Δ)RMS energy, (Δ)pitch, (Δ)HNR, (Δ)MFCC(1–12) | mean, standard deviation, kurtosis, skewness<br>extremes: min/max value, relative min/max, position, range<br>linear regression: offset, slope, MSE |

Apart from the pitch probability distribution features, we also use a 384 dimensions' feature set which is provided by INTERSPEECH Emotion Challenge 2009 [15] and is usually employed as a global feature set. The features are obtained by applying 12 functionals to several lowlevel descriptors (LLDs) including zero-crossing rate (ZCR), root mean square (RMS) energy, pitch, harmonics-to-noise ratio (HNR), and MFCC 1–12, together with their first order delta regression coefficients.

The whole LLDs and functionals in the feature set are shown in Table 1. These features are extracted automatically with the open resource toolkit openSMILE [16].

HNR, as one of the LLDs, is computed from the Autocorrelation Coefficient Function (ACF), and can be regarded as voicing probability. It is calculated as:

$$HNR(n) = 10 \log \frac{ACF(T_0)}{ACF(0) - ACF(T_0)} \quad (2)$$

$$ACF(\tau) = \sum_{m=0}^{N-1-\tau} x(m) x(m+\tau) \quad (3)$$

in which $N$, $x(m)$, and $T_0$ denote the fundamental period [17], the frame length, and the $m$ th sampling point in the $n$ th frame, respectively.

For the classification model, we used SVM with WEKA 3 Data Mining Toolkit [18]. Linear kernel is applied to avoid overfitting. Leave-one-out cross validation is performed for SVM training and testing to miximize the scale of training data.

In order to evaluate the effectiveness of our proposed method and features, several comparative experiments are conducted from different aspects. Firstly, experiment using global acoustic features of utterances without segmentation is regard as a benchmark. Then, experiments of different segmentation methods are conducted to verify the effectiveness of our proposed local dynamic pitch probability distribution features. In addition, principal components analysis (PCA) is employed for dimensionality reduction of the features.

### 3.2 Experimental results

In this paper, weighted average recall (WA, the number of correctly classified instances divided by the total amount of instances) and unweighted average recall (UA, the mean value of the recall for each class) are used to evaluate the performance of classification, in which the weighted average recall is able to reflect the overall accuracy for imbalanced class.

**Table 2.** Comparative results with different experiments in terms of (un-)weighted average recall (UA/WA)

| No. | Feature dimensions | WA (%) | UA (%) |
|---|---|---|---|
| 1 | 384 | 82.80 | 81.53 |
| 2 | 384+93=411 | 83.55 | 82.36 |
| 3 | 384*3=1152 | 84.11 | 83.77 |
| 4 | 384*3+93=1179 | 85.23 | 84.65 |
| 5 | 1179→PCA(99.0%)=409 | 85.42 | 84.87 |

Explanations:
- No. 1: Commonly-used global features of utterances without segmentation.

- No. 2: Pitch probability distribution features extracted from each segment together with commonly-used global features.
- No. 3: Commonly-used local features extracted from each segment.
- No. 4: Both pitch probability distribution features and commonly-used features extracted from each segment.
- No. 5: Apply PCA with a cumulative contribution rate of 99.0% to the features in No.4.

Table 2 presents the speech emotion recognition accuracies of different comparative experiments. Comparing the results of Experiment 1 without segmentation with the others, we find that time-based segmentation contributes to the accuracy with significant improvements. In addition, the pitch probability distribution features are able to increase the accuracy as well. Furthermore, with segmentation and the pitch probability distribution features applied together, the performance is further improved in Experiment 4. When the dimensionality is reduced to 409 with the utilization of PCA (cumulative contribution rate: 99.0%), the best result is achieved in Experiment 5, whose relative error rate is 18.08% lower than the benchmark in terms of UA. The improvement is achieved by the local dynamic information extracted with the segmentation approach.

**Table 3.** Confusion matrix for Experiment 1 (%)

|       | Hap.  | Neu.  | Ang.  | Sad.  | Fear  | Bore. | Dis.  |
|-------|-------|-------|-------|-------|-------|-------|-------|
| Hap.  | **73.24** | 1.41  | 16.9  | 0     | 5.63  | 0     | 2.82  |
| Neu.  | 2.53  | **84.81** | 2.53  | 0     | 1.27  | 8.86  | 0     |
| Ang.  | 7.87  | 0.79  | **90.55** | 0     | 0.79  | 0     | 0     |
| Sad.  | 0     | 1.61  | 0     | **80.65** | 0     | 16.13 | 1.61  |
| Fear  | 8.7   | 1.45  | 5.8   | 1.45  | **76.81** | 1.45  | 4.35  |
| Bore. | 0     | 2.47  | 0     | 9.88  | 1.23  | **86.42** | 0     |
| Dis.  | 4.35  | 2.17  | 2.17  | 0     | 8.7   | 4.35  | **78.26** |

**Table 4.** Confusion matrix for Experiment 5 (%)

|       | Hap.  | Neu.  | Ang.  | Sad.  | Fear  | Bore. | Dis.  |
|-------|-------|-------|-------|-------|-------|-------|-------|
| Hap.  | **73.24** | 2.82  | 16.9  | 0     | 7.04  | 0     | 0     |
| Neu.  | 2.53  | **91.14** | 1.27  | 0     | 1.27  | 3.8   | 0     |
| Ang.  | 11.02 | 0     | **88.19** | 0     | 0.79  | 0     | 0     |
| Sad.  | 0     | 3.23  | 0     | **85.48** | 1.61  | 9.68  | 0     |
| Fear  | 7.25  | 1.45  | 7.25  | 0     | **81.16** | 1.45  | 1.45  |
| Bore. | 0     | 6.17  | 0     | 2.47  | 1.23  | **90.12** | 0     |
| Dis.  | 0     | 4.35  | 2.17  | 4.35  | 0     | 4.35  | **84.78** |

The confusion matrices of the benchmark (Experiment 1) and the best result (Experiment 5) are given in Table 3 and 4. From the confusion matrices, we can observe that the performances of our proposed method in Experiment 5 are much better than which in Experiment 1 for most of emotions, which verified the effectiveness of our method.

Table 5 gives performances of proposed method in terms of UA on each emotion state. We observe that the segmentation and local dynamic pitch probability distribution features increase the performances of recognition of the majority of emotion states, except for happiness and anger. This result is understandable because happiness and anger utterances have similar dynamic trends on the pitch features [9] and therefore tend to be confused with each other.

**Table 5.** Effects of proposed method in terms of UA changes on each emotion state (%)

| Emotion | Benchmark | Best result | UA changes |
|---|---|---|---|
| Disgust | 78.26 | 84.78 | +6.52 |
| Neutral | 84.81 | 91.14 | +6.33 |
| Sadness | 80.65 | 85.48 | +4.84 |
| Fear | 76.81 | 81.16 | +4.35 |
| Boredom | 86.42 | 90.12 | +3.70 |
| Happiness | 73.24 | 73.24 | 0.00 |
| Anger | 90.55 | 88.19 | –2.36 |

**Table 6.** Results of different bins in the pitch histogram and number of segments under the experimental program of Experiment 5 in terms of UA (%)

| Interval in pitch histogram | 3 segments | 4 segments | 5 segments |
|---|---|---|---|
| 50 Hz | 84.87 | 83.16 | 81.78 |
| 25 Hz | 82.76 | 81.85 | 81.38 |

Besise, some further experiments are also conducted to explore how the number of segments and interval in the pitch histogram affect the recognition result. The experimental results show that the combination of pitch probability distribution features and commonly-used features extracted from each segment together with dimensionality reduction using PCA (i.e. the experimental program of Experiment 5) achieves the best result in each experimental condition. Table 6 shows the results under the experimental program of Experiment 5 in terms of UA (%). In order to examine the relationship between the number of the segments and the recognition results, experiments with four and five segments for each utterance are conducted, but the results are not as good as that with three segments. In addition, the UA decreases with the increase of the number of the segments. Moreover, the result with a bin width of 50 Hz is better than that of 25 Hz. A possible reason is that with smaller granularity of the segmentation and the pitch probability distribution features extraction, some of the emotional information is counteracted by the content differences in an utterance, and therefore it is adverse to the recognition of speech emotion.

## 4 Discussion and conclusion

In this paper, a novel local dynamic pitch probability distribution feature is proposed in time-based segments to improve the performance of speech emotion recognition. The experimental results suggest that the local dynamic information obtained by time-based segmentation and pitch probability distribution features are more effective for speech emotion recognition than those traditional global features. Some different segmentation related parameters are also examined in the experiments, the results show that too large or small granularity for the segmentation is adverse to the recognition of speech emotion.

There are several emotional speech corpora in various languages being used in the studies. The common problem, however, is that the scales of them are relatively small with respect to those for Automatic Speech Recognition (ASR), which usually makes it difficult to train the classifier well. Thus, it is also an issue to be addressed that how to achieve ideal performance with small-scaled training data. In addition, pitch is selected as one of the prosodic features that convey important information related to emotion for histogram calculation in this paper. Other features can also be analyzed in the similar way to expect a better performance in our future work.

In this study, we validate the dynamic nature of emotional speech in terms of features. Actually, the classification model influences the performance of recognition in large measure as well. Therefore, in the future, dynamic classification methods such as Recurrent Neural Network (RNN) will be considered since these sequential models are suitable for the dynamic information. Hybrid hierarchical models can also be attempted. Moreover, deep learning methods, which are based on the perception mechanism of the human brain, can be introduced for feature selection instead of traditional PCA method. Also, these features and approaches need to be evaluated on large-scaled dataset in order that the models can be trained enough.

## Acknowledgements

The research is supported partially by the National Basic Research Program of China (No. 2013CB329303), and the National Natural Science Foundation of China (No. 61303109 and No. 61503277). The study is supported partially by JSPS KAKENHI Grant (16K00297).